\documentstyle[12pt,psfig]{article}
\title{\bf Wolf-Rayet Stars in Very Young Starburst Galaxies
\footnote{Based on data collected at the European Southern Observatory (La
Silla, Chile)} }
\author{Thierry Contini
\vspace{1cm}\\
\normalsize Laboratoire d'Astrophysique (UMR 5572), Observatoire
Midi-Pyr\'en\'ees, France
}
\date{\mbox{}}
\begin{document}
\maketitle
\pagestyle{empty}
%
%
\def\bull{\vrule height .9ex width .8ex depth -.1ex}
\makeatletter
\def\ps@plain{\let\@mkboth\gobbletwo
\def\@oddhead{}\def\@oddfoot{\hfil\tiny\bull\quad
``WR Stars in the Framework of Stellar Evolution'';
33$^{\mbox{\rm rd}}$ Li\`ege\ Int.\ Astroph.\ Coll., 1996\quad\bull}%
\def\@evenhead{}\let\@evenfoot\@oddfoot}
\makeatother
%
%
\def\beginrefer{\section*{References}%
\begin{quotation}\mbox{}\par}
\def\refer#1\par{{\setlength{\parindent}{-\leftmargin}\indent#1\par}}
\def\endrefer{\end{quotation}}
%
%
{\noindent\small{\bf Abstract:} 
Preliminary results from spectrophotometric observations of galaxies
with very young starbursts are presented. Starburst galaxies with an age 
of the burst in the range between 3 and 6 Myr have been observed and new 
detections of Wolf-Rayet galaxies are reported. We discuss the origin of 
high excitation nebular lines observed in these galaxies and their possible 
link with the population of Wolf-Rayet stars .
}     
%
%
\section{Introduction}

To derive the stellar population in starburst galaxies is one of the fundamental goals in studies of star forming regions. Because of their 
large luminosities, massive stars and especially Wolf-Rayet (W-R) stars 
are directly visible in the integrated spectrum of young starbursts and thereby provide an unique opportunity to study their stellar content.

In a subset of starburst galaxies, often referred to as Wolf-Rayet galaxies (Conti 1991), {\em broad stellar} emission lines (He\,{\sc ii}$\lambda$4686 for the strongest) in their optical spectrum testify to an important population of W-R stars (Kunth \& Sargent 1981, Kunth \& Schild 1986). The very high ratio of W-R to O stars in W-R galaxies 
(Vacca \& Conti 1992, Contini et al.~1995) is compatible with models of stellar evolution computed by Maeder \& Meynet (1994), provided that star formation occurs in {\it burst} mode on very short time scales, less than 10$^6$ years, and with a rather flat IMF (Contini et al.~1995).

It is now essential to enlarge the sample of known W-R galaxies. The detailed
study of this type of galaxies is of considerable importance for models of
stellar evolution and of starbursts. Indeed, because W-R stars are the direct
offsprings of the most massive O stars, and because their lifetime is at most
10$^6$ years, the presence of a very large number of such stars in a galaxy
provides important constraints on several parameters characterizing starbursts, such as duration, intensity and age of the burst, and, most important, its IMF.

There is certainly a large number of W-R galaxies in the Universe, but only 
about 70 such galaxies have been inventoried so far, with about 40 objects 
reported in the catalog of Conti (1991). Since the first discovery by Allen et al.~(1976), most of these galaxies have been discovered serendipitously and only two systematic searchs have been realized with limited success (Kunth \& Joubert 1985, Masegosa et al.~1991). Because of the high sensitivity of CCDs mounted on 2-meter class telescopes, a systematical search for W-R galaxies can now be done. 

This paper presents preliminary results of spectrophotometric observations 
of galaxies with very young starbursts. Using observational data (Terlevich 
et al.~1991, Contini 1996) and starbursts models (Cervi\~no \& Mas-Hesse 
1994, Leitherer \& Heckman 1995), we selected starburst galaxies with an age 
of the burst in the range between 3 and 6 Myr, which corresponds to the 
predicted Wolf-Rayet phase in starburst galaxies (Meynet 1995, Schaerer 1996). 

\section{Sample Selection}

\begin{table}
\begin{center}
{\small
\begin{tabular}[c]{|l||ccrrcc|}	
\hline
Name & $\alpha$ & $\delta$ & V$_{\rm rad}$ & W(H$\beta$) & (O/H) & Starburst age \\
  &   &  & kms$^{-1}$ & \AA\ & $\odot$ & Myr \\
\hline
ESO 566-8 & 09$^h$ 42$^m$ 39$^s$ & --19$^{\circ}$ 28' 54" & 9750 & 60 & &  \\
Mkn 712 & 09$^h$ 53$^m$ 59$^s$ & +15$^{\circ}$ 52' 34" & 4550 & 63 & 0.30 & \hfil 3 - 4  \\
Tol 4 & 10$^h$ 08$^m$ 00$^s$ & --28$^{\circ}$ 42' 00" & 4197 & 125 & 0.15 & \hfil 3 - 3.5 \\
Mkn 1271 & 10$^h$ 53$^m$ 33$^s$ & +06$^{\circ}$ 26' 24" & 1027 & 104 & 0.18 & \hfil 3 - 4 \\
UM 462 - A & 11$^h$ 50$^m$ 04$^s$ & --02$^{\circ}$ 11' 28" & 986 & 69 & 0.22 & \hfil 4 - 4.5 \\
UM 462 - B & 11$^h$ 50$^m$ 04$^s$ & --02$^{\circ}$ 11' 28" & 986 & 124 & 0.19 & \hfil 3 - 3.5 \\
UM 469 & 11$^h$ 54$^m$ 39$^s$ & +02$^{\circ}$ 45' 10" & 17388 & 68 & 0.26 & \hfil 4 - 4.5 \\
Mkn 1318 & 12$^h$ 16$^m$ 36$^s$ & +04$^{\circ}$ 08' 07" & 1527 & 121 & 0.26 & \hfil 3 - 3.5 \\
Tol 30 & 13$^h$ 03$^m$ 03$^s$ & --28$^{\circ}$ 09' 12" & 2099 & 161 & 0.20 & \hfil 3 - 3.5 \\
ESO 513-11 - A & 14$^h$ 57$^m$ 31$^s$ & --26$^{\circ}$ 15' 06" & 5115 & 78 & 0.18 & \hfil 3 - 4 \\
ESO 513-11 - B & 14$^h$ 57$^m$ 31$^s$ &  --26$^{\circ}$ 15' 06" & 5115 & 84 & 0.21 & \hfil 3 - 4 \\
\hline
\end{tabular} 
}
\end{center}
\caption{Sample of galaxies with young starbursts (age between 3 and 6 Myr). The starburst ages were estimated by comparison of spectrophotometric data (equivalent width of H$\beta$ and oxygen abundance) with predictions of the evolutionary population synthesis models of Cervi\~no \& Mas-Hesse (1994) and Leitherer \& Heckman (1995).}
\end{table}

The sample of W-R galaxy candidates was extracted from the spectrophotometric catalogs of H\,{\sc ii} galaxies (Terlevich et al.~1991) and of barred starburst galaxies (Contini 1996). 
The age of starbursts in these galaxies was estimated by comparison of
spectrophotometric data (equivalent width of H$\beta$ and oxygen abundance) with predictions of evolutionary population synthesis models of Cervi\~no \& Mas-Hesse (1994) and Leitherer \& Heckman (1995).


We computed the oxygen abundance (O/H) of all galaxies by using the dereddened fluxes of the oxygen and H$\beta$ emission lines. The oxygen abundance was derived using the [O\,{\sc iii}]$\lambda\lambda$4959,\-5007 emission lines for the catalog of Contini (1996). For the catalog of Terlevich et al.~(1991) we also used the [O\,{\sc ii}]$\lambda$3727 line. These values, which are used as an indicator of the metallicity of H\,{\sc ii} regions, were determined from the empirical relationship between log(O/H) and the intensities of the [O\,{\sc iii}] and [O\,{\sc ii}] lines given by Edmunds \& Pagel (1984) with a linear fit given by Vacca \& Conti (1992). 

To derive with a good accuracy ($\pm$ 0.5 Myr) the age of starbursts, we used the H$\beta$ equivalent width for a given metallicity as the
observational parameter. These values are then compared to predictions of two recent starburst models of Cervi\~no \& Mas-Hesse (1994) and
Leitherer \& Heckman (1995). We then selected starburst galaxies with an estimated age of the burst between 3 and 6 Myr, which corresponds to the predicted W-R phase in starburst galaxies (Meynet 1995, Schaerer 1996). A list of W-R galaxy candidates with positive detections is given in Table 1 where the spectrophotometric data (equivalent width of H$\beta$ and oxygen abundance) and the estimation of starburst age are reported.

\section{Observations and Data Reduction}

Long-slit spectroscopic observations of the galaxies were obtained on the nights of 1996 March 15 -- 20 at the 1.52-meter telescope of ESO. The data were acquired with the Boller \& Chivens spectrograph and a 2048$\times$512 Ford Aerospace CCD. The spectral resolution was 180 \AA/mm, and provided a spectral coverage of 3100 -- 8700 \AA\ with a resolution $\Delta \lambda \sim$ 10 \AA\ as mesured from the FWHM of gaussian profiles adjusted to the He-Ar comparison lines. 

The slit width ($\sim$ 3") and positions were chosen to cover entirely the brightest H\,{\sc ii} regions of the galaxies. During the observing run the mean spatial resolution was about 1 to 1.5 arcsec. In order to accurately detect
low intensity emission lines in the W-R bump, the exposure time was between 1 and 2 hours to reach S/N ratios greater than 10. The time integration for each galaxy was divided into exposures of about 20 minutes in order to avoid saturation of the bright nebular lines ([O\,{\sc iii}] and H$\alpha$) and to recognize cosmic impacts. During the run we also observed the spectrophotometric standard stars GD 108 and G 60-54 taken from the list given by Oke (1990) with a slit width of about 7 " in order to flux calibrate the spectrum of the galaxies.

The spectroscopic data were reduced according to a standard reduction procedure using the MIDAS package LONG. These include bias subtraction, flat-field corrections, wavelength calibration, sky subtraction and cosmic ray removal.

The two-dimensional spectra of the galaxies were then corrected for foreground extinction in our own Galaxy. We accounted for the foreground reddening using the values of the total galactic extinction E(B-V) given by Burstein \& Heiles (1984) and the average Galactic extinction law of Howarth (1983) with $R_V$ = 3.1, covering entirely the spectral range of interest. 

Finally, we extracted one-dimensional elementary spectra by averaging several columns (spatial dimension). The elementary spectra correspond to each H\,{\sc ii} region of the galaxies and are used to enhance the S/N ratio for the detection of weak emission lines in the W-R bump. The spectra were averaged over 3 arcsec which corresponds to the mean projected size of H\,{\sc ii} regions at the distance of the observed galaxies (see Table 1).

\section{Results and Discussion}

\begin{table}
\begin{center}
{\small
\begin{tabular}[c]{|l||ccccc|}	
\hline
Name & N\,{\sc iii} & [Fe\,{\sc iii}] & He\,{\sc ii} & [Ar\,{\sc iv}] & [Ar\,{\sc iv}] \\
  & 4640 \AA\ & 4658 \AA\ & 4686 \AA\ & 4711 \AA\ & 4740 \AA\ \\
\hline
ESO 566-8 &  & X & narrow &  & \\
Mkn 712 & {\bf broad} & X & {\bf broad} & & \\
Tol 4 &  & & narrow & X & X \\
Mkn 1271 & & X & narrow & X & X  \\
UM 462 - A & & X & narrow & X & \\
UM 462 - B & & X & narrow & X & X   \\
UM 469 &  & X & narrow & X & \\
Mkn 1318 & & X & narrow & X & X \\
Tol 30 & {\bf broad} & & {\bf broad} & X & X \\
ESO 513-11 - A &  & X & {\bf broad} & X &   \\
ESO 513-11 - B &  & X & narrow &  &    \\
\hline
\end{tabular} 
}
\end{center}
\caption{Result for the detection of W-R stars in the integrated spectrum of galaxies. X : detection of a narrow (FWHM $\sim$ 10 \AA) nebular emission line of highly 
ionized species. {\bf broad} : broad (FWHM $\sim$ 30 \AA) stellar emission line from W-R stars. {\em narrow} : only a narrow nebular He\,{\sc ii}$\lambda$4686 emission is detected.}
\end{table}

The main signature of Wolf-Rayet stars in galaxies beyond the Local Group is the presence of ``broad" emission lines (He\,{\sc ii}$\lambda$4686, N\,{\sc iii}$\lambda$4640 for the strongest) in their integrated optical spectrum. We then have to search for these spectral signatures in the spectrum of galaxies with young starbursts. The results of detection are summarized in Table 2.

We detect broad emission lines (FWHM $\sim$ 30 \AA) in three galaxies of the sample, namely Mkn 712, Tol 30 and ESO 513-11. In the two first galaxies (Figs. 1 and 2), one broad emission line is clearly identified with He\,{\sc ii}$\lambda$4686, the other is uncertain, corresponding to either N\,{\sc iii}$\lambda$4640 or C\,{\sc iii}/{\sc iv}$\lambda\lambda$4650,4658. The difficulty of identifying this second broad emission line is mainly due to our low spectral resolution. In the third galaxy ESO 513-11, we only detect the broad He\,{\sc ii}$\lambda$4686 emission line. The same type of spectral signature is present in the seventy or so known W-R galaxies (Conti 1991, Vacca \& Conti 1992) and attributed to the presence of a large number of W-R stars. {\em Tol 30 and ESO 513-11 can then be classified as new Wolf-Rayet galaxies whereas Mkn 712 was already known} (Contini et al.~1995).

In addition to these broad emission features, we detected with a high confidence level three narrower (FWHM $\sim$ 10 \AA) nebular lines, namely 
[Fe\,{\sc iii}]$\lambda$4658 and [Ar\,{\sc iv}]$\lambda\lambda$4711,4740 (see Table 2). All galaxies of the sample exhibit the narrow nebular He\,{\sc ii}$\lambda$4686 emission line (see Figs 3 and 4). In these conditions, a nebular contribution to the He\,{\sc ii}$\lambda$4686 emission line is expected in the three detected W-R galaxies Mkn 712, Tol 30 and ESO 513-11. This is supported by Figs. 1 and 2 where the He\,{\sc ii}$\lambda$4686 feature appears to be composed of a narrow component atop a broader base. 

\medskip
While the detection of broad emission lines in the spectrum of starburst
galaxies is clearly associated with W-R stars, why are there also high 
excitation {\em nebular} lines ? 
\medskip

The origin of the narrow He\,{\sc ii}$\lambda$4686 line in starburst galaxy spectra is not yet completely understood, since its excitation requires a relatively hard ionizing spectrum ; for a stellar ionizing source, an effective temperature of $T_{eff} \geq$ 70 000 K would be needed (Garnett et al.~1991). Standard models of O stars spectra indicate that even at these temperatures very few photons with energies $E \geq$ 54 eV are produced. However, based on recent NLTE stellar atmosphere models Gabler et al.~(1992) argued that O stars close to the Eddington limit might produce enough high energy photons. Motch et al.~(1993) suggested that hot WN stars could be the origin of the line. Alternatively, Garnett et al.~(1991) suggested that this line could also be produced by shocks or massive X-ray binaries. 

If W-R stars are responsible for the nebular He\,{\sc ii}$\lambda$4686, W-R features should be evident in the optical spectra of these galaxies. But many galaxies exhibit only {\it nebular} (as opposed to {\it stellar}) He\,{\sc ii}$\lambda$4686 emission line. This probably means that, if these galaxies do have W-R stars, the latter must be in very small numbers. 

Schaerer (1996) considers that the relatively hard spectra of WC stars may be responsible for the origin of nebular He\,{\sc ii}$\lambda$4686 emission. Using his latest CoStar model (Schaerer et al.~1996a, 1996b) fluxes for O stars and appropriate atmosphere models for W-R stars (Schmutz et al.~1992), he shows that the major He\,{\sc ii} nebular emission is due to stars in the WC phase for a age of the burst between 3 and 5 Myr. This contribution is particularly important for metallicities between 0.2 Z$_{\odot}$ and Z$_{\odot}$ which correspond to the metallicity estimated in the galaxies observed. WC stars have  negligible broad stellar emission in the He\,{\sc ii}$\lambda$4686 line ; this may explain the non detection of broad stellar line in the spectrum of the majority of galaxies observed.
 
To test the various suggestions for the origin of the nebular He\,{\sc ii}$\lambda$4686 line and determine if there is a link between the population of massive star (W-R and/or O stars) and the nebular He\,{\sc ii}$\lambda$4686 emission, it is  now crucial to measure the relative contributions of the {\em broad stellar} and {\em narrow nebular} emission in the He\,{\sc ii}$\lambda$4686 line, and to search for weak stellar emission features (especially C\,{\sc iv}$\lambda$5808 for WC stars) in those galaxies for which nebular He\,{\sc ii}$\lambda$4686 emission has been detected but W-R signatures have not been seen.    

%
%
\section*{Acknowledgements}
I would like to thank Emmanuel Davoust for carefuly reading this manuscript and for his comments. I am grateful to Daniel Schaerer and Bill Vacca for stimulating discussions during this conference. 
%
%
 
\beginrefer
\refer Allen D.A., Wright A.E., Goss W.M., 1976, MNRAS 177, 91

\refer Burstein D., Heiles C., 1984, ApJS 54, 33

\refer Cervi\~no M., Mas-Hesse J.M., 1994, A\&A 284, 749

\refer Conti P.S., 1991, ApJ 377, 115

\refer Contini T., 1996, PhD Thesis, University of Toulouse, France

\refer Contini T., Davoust E., Consid\`ere S., 1995, A\&A 303, 440
 
\refer Edmunds M.G., Pagel B.E.J., 1984, MNRAS 211, 507

\refer Gabler R., Gabler A., Kudritzki R.P., Mendez R.H., 1992, A\&A 265, 656

\refer Garnett D.R., Kennicutt R.C., Chu Y., Skillman E.D., 1991, ApJ 373, 458

\refer Howarth I.D., 1983, MNRAS 203, 301

\refer Leitherer C., Heckman T.M., 1995, ApJS 96, 9

\refer Kunth D., Joubert M., 1985, A\&A 142, 411

\refer Kunth D., Sargent W.L.W., 1981, A\&A 101, L5

\refer Kunth D., Schild H., 1986, A\&A 169, 71


\refer Maeder A., Meynet G., 1994, A\&A 287, 803

\refer Masegosa J., Moles M., del Olmo A., 1991, A\&A 244, 273

\refer Meynet G., 1995, A\&A 298, 767

\refer Motch C., Pakull M.W., Pietsch W., 1993. In: Tenorio-Tagle G. (eds.), Violent 
Star Formation : From 30 Doradus to QSO's. Cambridge University Press, Cambridge, 
p.~208

\refer Oke J., 1990, AJ 99, 1621


\refer Schaerer D., 1996, ApJ 467, L17

\refer Schaerer D., de Koter A., Schmutz W., Maeder A., 1996a, A\&A, in press

\refer Schaerer D., de Koter A., Schmutz W., Maeder A., 1996b, A\&A, in press

\refer Schmutz W., Leitherer C., Gruenwald R., 1992, PASP 104, 1164

\refer Terlevich R., Melnick J., Masegosa J., Moles M., Copetti M.V.F., 1991, A\&AS
91, 285

\refer Vacca W.D., Conti P.S., 1992, ApJ 401, 543

\endrefer           

\begin{figure}
\psrotatefirst
\centerline{\psfig{file=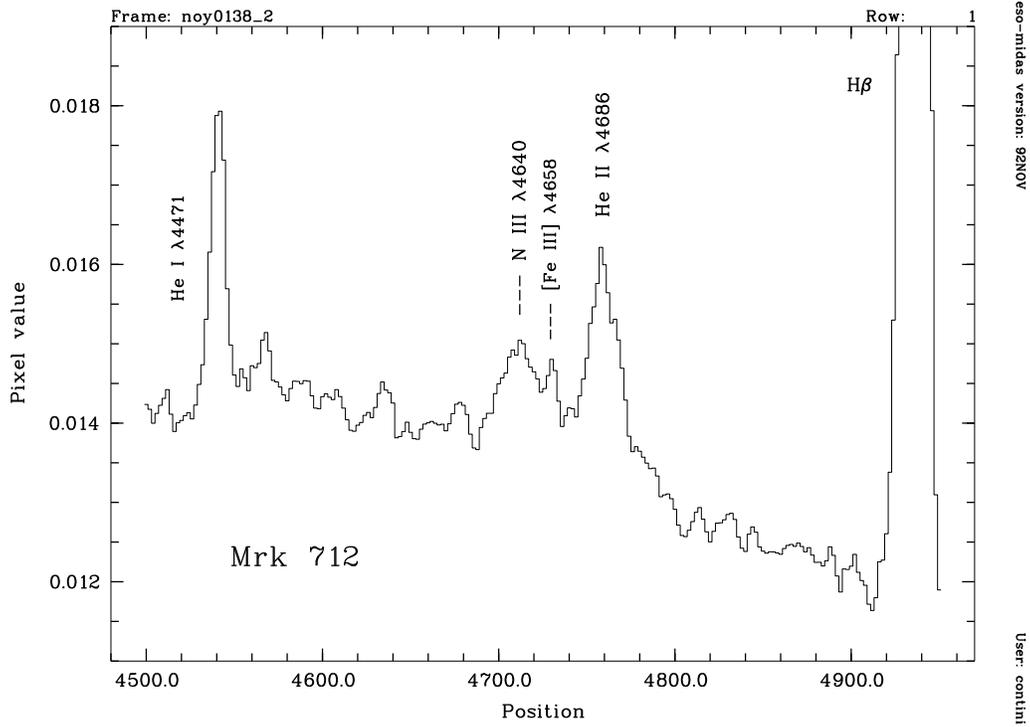,height=10cm,angle=-90}}
\caption{The Wolf-Rayet ``bump" in the dereddened optical spectrum of the Wolf-Rayet galaxy Mkn 712. The He\,{\sc ii}$\lambda$4686 emission line consists of a narrow component due to nebular emission atop a broader component due to emission from Wolf-Rayet stars. Intensities are in units of 10$^{-14}$ erg s$^{-1}$ cm$^{-2}$ \AA$^{-1}$. Wavelengths are expressed in \AA.}
\end{figure}

\begin{figure}
\psrotatefirst
\centerline{\psfig{file=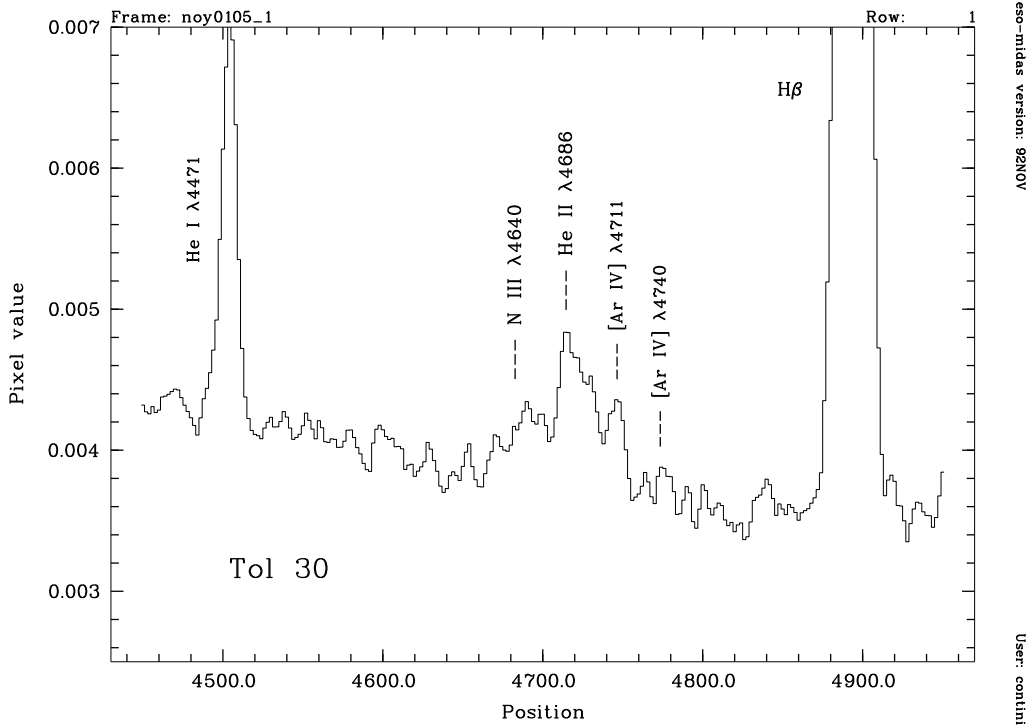,height=10cm,angle=-90}}
\caption{The Wolf-Rayet ``bump" in the dereddened optical spectrum of the new Wolf-Rayet galaxy Tol 30. See Fig. 1 for comments on units.}
\end{figure}

\begin{figure}
\psrotatefirst
\centerline{\psfig{file=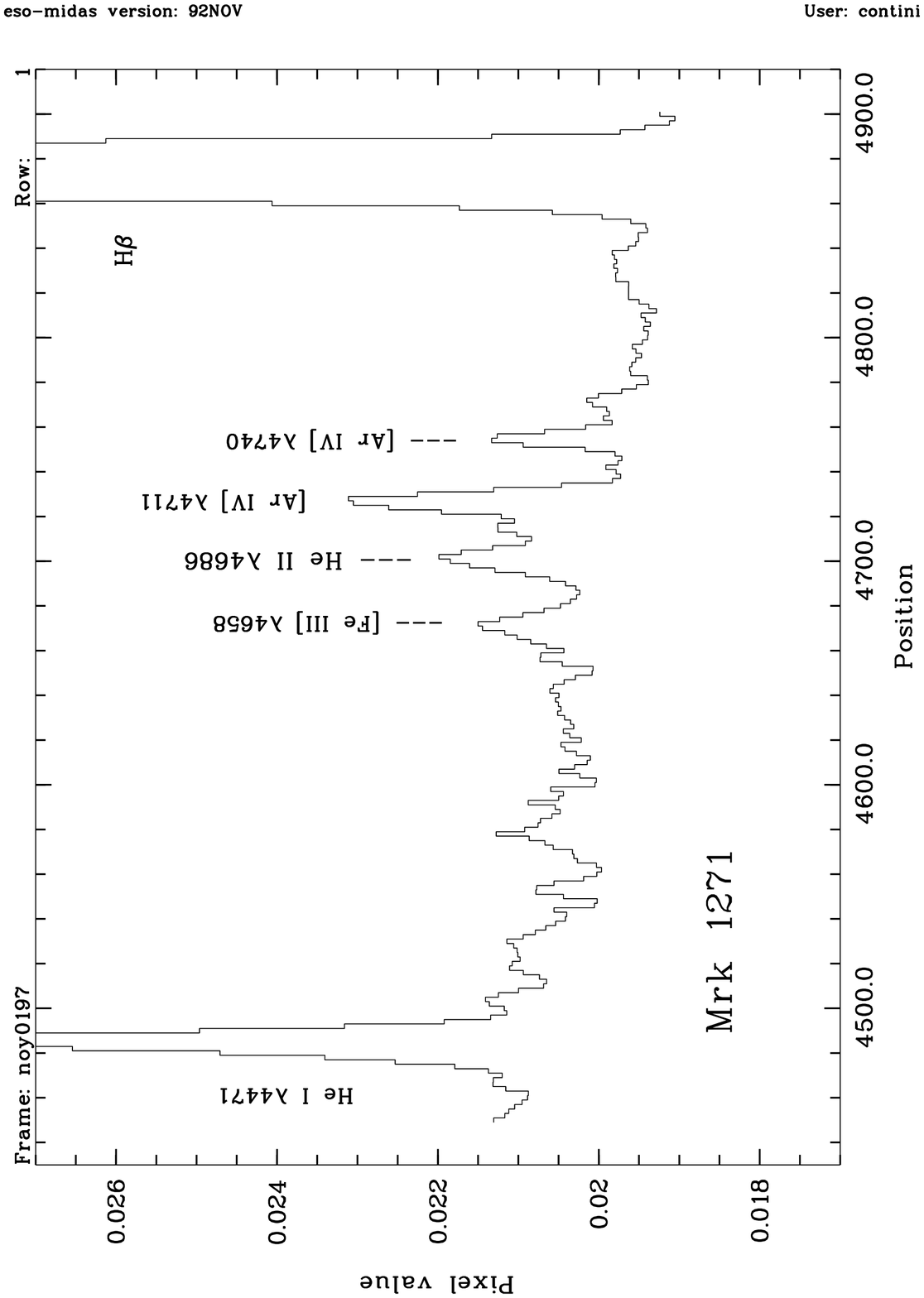,height=10cm,angle=-90}}
\caption{Set of nebular emission lines of highly ionized species ([Fe\,{\sc iii}], He\,{\sc ii} and [Ar\,{\sc iv}]) in the optical spectrum of Mkn 1271. 
See Fig. 1 for comments on units.}
\end{figure}

\begin{figure}
\psrotatefirst
\centerline{\psfig{file=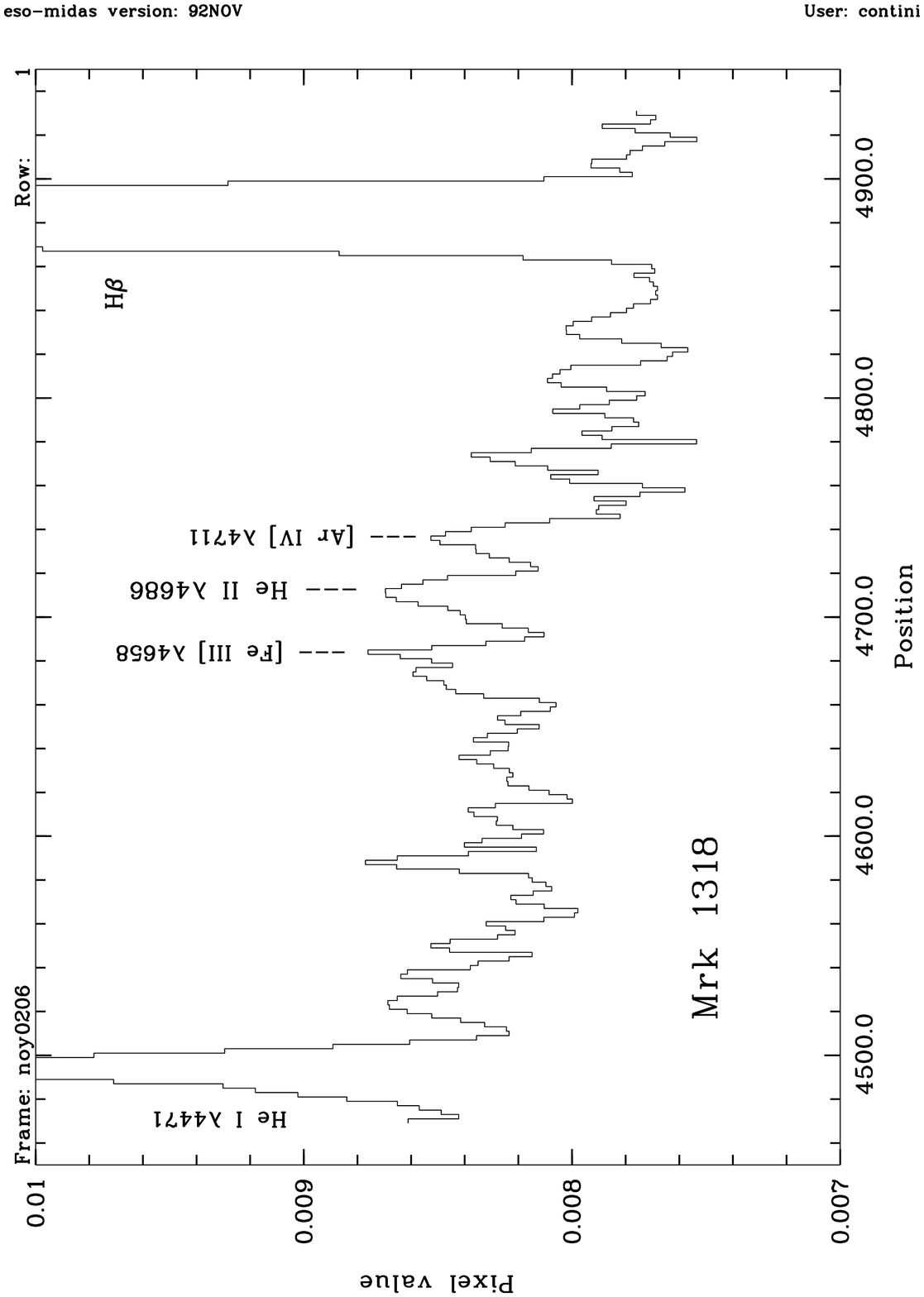,height=10cm,angle=-90}}
\caption{Set of nebular emission lines of highly ionized species ([Fe\,{\sc iii}], He\,{\sc ii} and [Ar\,{\sc iv}]) in the optical spectrum of Mkn 1318. 
See Fig. 1 for comments on units.}
\end{figure}

\end{document}